# All-dielectric self-cloaked structures

Zeki Hayran,[1,*] Ramon Herrero,[2] Muriel Botey,[2] Hamza Kurt,[1] and Kestutis Staliunas[2,3]

[1] *Nanophotonics Research Laboratory, TOBB University of Economics and Technology, Department of Electrical and Electronics Engineering, Ankara, Turkey*
[2] *Departament de Física, Universitat Politècnica de Catalunya (UPC), Barcelona, Spain*
[3] *Institució Catalana de Recercai Estudis Avançats (ICREA), Barcelona, Spain*
*Corresponding author: zekihayran@etu.edu.tr*

**While practical realizations of optical invisibility have been achieved so far by various ingenious methods, they generally rely on complex materials which prevent the wide implementation of such schemes. Here, we propose an alternative indivisibility procedure to design objects (*i.e.* self-cloaked structures) that have optical properties identical to the surrounding environment and are, thereby, intrinsically invisible to an external observer as such (without the necessity of an external cloak). The proposed method is based on the uncoupling of the scattered waves from the incident radiation by judiciously manipulating the scattering potential of a given object. We show that such a procedure is able to yield optical invisibility for any arbitrarily shaped object within any specified frequency bandwidth by simply employing isotropic non-magnetic dielectric materials, without the usage of loss or gain material. The validity of the design principle has been verified by direct experimental observations of the spatial electric field profiles and scattering patterns at the microwave regime. Our alternative self-cloaking strategy may have profound implications especially in noninvasive probing and cloaked sensor applications, where the wave penetrability into the sensor region is essential together with its invisibility to minimize the field distortion.**

The deflection of a light ray from a straight path, known as light scattering, is one of the most common interactions between light and matter. As light impinges on an irregularity in space (e.g. object, particles, interface between two different media) it will experience partial reflection and transmission. While such phenomenon is responsible for creating marvelous light events in everyday life, such as the iridescent colors of butterfly wings, it may be desirable to cancel it for some applications. The so-called invisibility cloaking effect is a recent thought-provoking discovery, where the optical properties of the background and/or the object are tailored such that the light propagation mimics that of the free space and, thus, the object is effectively invisible to an external observer. A major breakthrough in 2006 [1,2], now known as transformation optics, has been achieved in this direction by deriving the spatial refractive index variation from the associated coordinate transformations required for a specific wave operation. Typically, to enable a cloaking effect with such a procedure, the physical space has to be distorted so that the incoming wave is rerouted around the object that is to be cloaked [3-5]. In effect, the distortion of the isotropic space results in an anisotropic and inhomogeneous index profile. Hence, due to the experimental burden of such a profile, various simplification attempts have been made to mitigate the fabrication requirements [6-9], which, however, inevitably result in scarifying of the ideal cloaking performance.

Apart from transformation optics based cloaking design procedures, a technique called scattering cancellation may also be employed to render a particular object invisible. In this technique, a properly designed cover layer results in wave scattering that effectively cancel the scattering of the object [10]. In this regard, for instance plasmonic covers [11], multi-layered dielectric coatings [12] and acoustic surface impedances [13] have been shown to significantly reduce the scattering effect of an object.

Another completely different method to suppress the back-scattered waves [14] and provide invisibility [15] has been recently proposed by employing the well-known Kramers-Kronig relations to relate the real and imaginary parts of a complex dielectric susceptibility profile. As such structures are already invisible *per se*, they do not require any external cloak and are a promising tool especially for cloaked sensor applications [16]. However, these methods suffer from various drawbacks. Firstly, as the susceptibility profile needs to be necessarily complex such that loss [14] or even gain [15] materials are required for a practical realization, which are often not easy to tailor in a bulk material. Secondly, the resultant susceptibility profile that is rendered invisible consists of an interface with infinitely extended tails rather than a specific object with a given shape and size. To form arbitrarily shaped self-cloaked objects, another study [17] suggested the usage of corrugated metallic wires. However, since the reported method relies on the resonant nature of the metallic wires, the self-cloaking effect has been demonstrated only for a narrow frequency bandwidth. Hence, broadband self-cloaked structures are yet to be elucidated and practically realized.

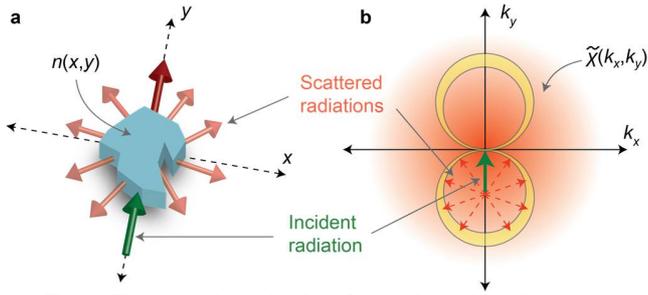

**Fig. 1.** The scattering behavior of (a) a given object is determined by (b) its scattering potential and, thus, by a judicious inverse-design of the scattering potential one can eliminate undesired scattering directions (indicated by red dashed lines). For the cloaking effect all scattering directions except the incident direction are eliminated.

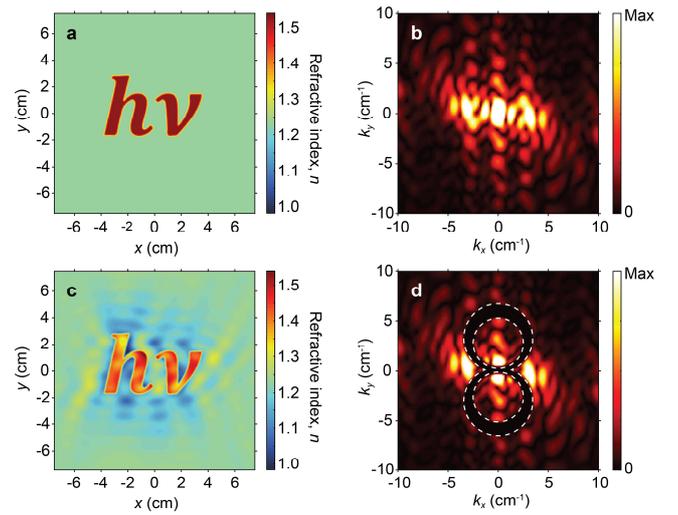

**Fig. 2.** The refractive index distribution of (a) the initial object (exhibiting no cloaking effect) and (b) its associated scattering potential are given. After applying the cloaking procedure, the resultant cloaked refractive index profile and its scattering potential are depicted in (c) and (d), respectively. The white dashed lines in (d) outline the filtered region in wavevector space. The background index is equal to 1.23.

In this manuscript, we present a unique approach to achieve a self-cloaking effect (thus not requiring any external cloak) for an arbitrary object within a specified frequency bandwidth and in particular propagation directions. We show how to obtain the required spatial index profile, which consists only of isotropic dielectric materials, that yields a scattering cancellation effect such that the object becomes cloaked, creating the illusion of restoring the original wavefront after leaving the object. We demonstrate this effect experimentally at the microwave regime by observing the spatial field profiles and by further investigating the frequency resolved scattering patterns.

The main idea of our proposal is based on the fact that the scattering behavior of a given object, or more generally a spatial dielectric susceptibility distribution, is associated with its spatial Fourier spectrum. We recently demonstrated and provided a generalized Hilbert transform on how this property can be exploited to tailor the scattering response of a given arbitrary object on a specific demand [18]. Here, to generate a cloaking effect and render a given object cloaked to the impinging electromagnetic radiation, all scatterings except in the incident direction should be cancelled. Such a Fourier interpretation is briefly illustrated in Fig. 1. Figure 1(a) depicts the spatial refractive index distribution $n(x,y)$ of a given arbitrary object, whereas Fig. 1(b) representatively shows the associated scattering potential $\tilde{\chi}(k_x, k_y)$, where $k_x$ and $k_y$ are the wavevectors along the $x$- and $y$- directions, respectively. The scattering potential $\tilde{\chi}(k_x, k_y)$ can be defined as the two-dimensional spatial Fourier transform of the dielectric susceptibility $\chi(x,y) = n^2(x,y) - 1$. As mentioned above, to eliminate all backward and forward scattered waves at a specific frequency, the Fourier components along a constant wavevector circle centered at the corresponding operational wavevector should be eliminated. On the other hand, to cancel the scattered waves for a specific frequency interval, the wavevector area (yellow colored areas in Fig. 1(b)) between the corresponding two constant wavevector circles should be removed, as shown in Fig. 1(b). Interesting to point out is that for a bidirectional cloaking effect, two symmetrically placed areas between constant wavevector circles should be removed. Since such a symmetry condition implies a purely real susceptibility when inverse Fourier transforming the scattering potential, the cloaked object would consist only of real refractive indices. In other words, no gain or loss materials would be required. Here, also important to note is that even though the presented method in Fig. 1 relies on the Born approximation [14] (which requires very weak scattering and, hence, a low index contrast), we expect the operational principle to work also under large index differences. This is in particular due to fact that the spatial wavevector would not undergo a significant change when interacting with a scatterer that is designed to be cloaked (as the wave would propagate as in free space without any change in its spatial wavelength). Hence, the locations of the operational constant wavevector circles at the wavevector domain would remain almost constant and, in result, the Born approximation would be a good assumption even for scatterers with moderately large index contrasts, as will be demonstrated below.

To exemplify the proposed principle, we start with an object shaped in the letters "*hv*" whose refractive index profile $n(x,y)$ and spatial dimensions are shown in Fig. 2(a). The scattering potential $\tilde{\chi}(k_x, k_y)$ of such an object is given in Fig. 2(b), which has been calculated as the Fourier transform of its dielectric susceptibility. Here, for demonstration purposes, we specify the positions of the inner and outer constant wavevector circles as $(k_x, k_y) = (0, \pm 2.60)$ cm$^{-1}$ and $(0, \pm 3.10)$ cm$^{-1}$, respectively; whereas their radii are equal to 2.50 cm$^{-1}$ and 3.20 cm$^{-1}$, respectively; which result in a bidirectional cloaking effect in both the $+y$ and the $-y$ directions within the frequency interval of 10.5 GHz and 12.9 GHz. The position and the wavevector circles have been determined so that the cloaking frequency interval and the material index range are within the ranges of our experimental equipment. Moreover, as opposed to Fig. 1(b), the wavevector region has a non-zero thickness at the center to efficiently eliminate the forward scattered waves. By applying the abovementioned procedure, the spatial index distribution and the scattering potential would be obtained as in Figs. 2(c) and 2(d), respectively. Important to note is that the index variation (and hence the index of the background) depends on the material to be used, similarly as in transformation optics based cloaking devices [19]. In the rest of this study, we have employed clear cast acrylic sheets to realize the original and cloaked objects to operate at the microwave regime. In this regard, the index variation presented in Fig. 2 has been adjusted (by adding a constant index throughout the medium) to fit inside the index range realizable by perforated acrylic. Using such a material we have fabricated three structures: a reference structure consisting only of a background index (1.23), a structure with the original object (as in Fig. 2(a)) and a structure with the cloaked object

(as in Fig. 2(c)). For a detailed description of the effective index calculation and the fabrication procedure see Supplement 1, Section 1. The experimental setup and the fabricated structures used to characterize the cloaking performance are shown in Figs. 3(a) and 3(b), respectively. As can be seen from Fig. 3(a), the experimental setup consists of a vector network analyzer (Agilent E5071C ENA), a motorized linear stage, a transmitter standard gain horn antenna and a coaxial monopole antenna (for details of the measurement technique please refer to the Supplement 1, Section 2). The obtained spatial field distributions at the operational frequency 12.0 GHz are collectively given in Fig. 4 for the +y and the –y directions, along with the corresponding numerical simulations. The numerical calculations have been conducted via the finite-difference time-domain (FDTD) method [20] in a three-dimensional grid with perfectly matched layers at the computational boundaries. It follows from these figures that the original object (Figs. 4(b), 4(e), 4(h) and 4(k)) causes light scattering and wave deformation, whereas the field propagation along the designed cloaked object (Figs. 4(c), 4(f), 4(i) and 4(l)) mimics well that of the reference structures (Figs. 4(a), 4(d), 4(g) and 4(j)). We should note that the cloaking performance is not perfect, which we mainly attribute to the discretization of the continuous index distribution. Moreover, another point worth mentioning is that the presented design principle assumes an incident wavevector in the +y or the –y direction, while the transmitter horn antenna radiates electromagnetic waves with spherical wavefronts. This might seem incongruous at first, however one should notice that the extended area of the filtered wavevector region (see Fig. 2(d)) may also enable invisibility for grazing incidence angles. The reason for this is that in the case of positioning the constant wavevector circle at the regarding operational frequency with a grazing angle offset, the constant wavevector circle would still overlap with the filtered wavevector region and, thus, the scattered waves would be eliminated. We may attribute the slight discrepancy between numerical and experimental results in Fig. 4 to the coupling effect of the receiver antenna's metallic tip with the electromagnetic field along the structure. Such a coupling effect potentially distorts the field measurement especially when the detector is close to the structure. Besides, the evanescent tails of the propagating mode inside the structure decrease for a higher refractive index. Hence, in comparably higher refractive index regions the experimentally detected field amplitude is lower. This is especially noticeable around the center region (where the object is placed), see Fig. 4, where the detected field amplitude is comparably lower precisely due to the high refractive index of the object. Furthermore, the field distributions for the +x and –x directions, for which the object does not exhibit invisibility, have also been measured and provided in Fig. S2 (Supplement 1, Section 3). In addition, measured field distributions at the operational frequencies 10.5 GHz and 12.9 GHz have been provided in Figs. S3 and S4, respectively (Supplement 1, Section 3).

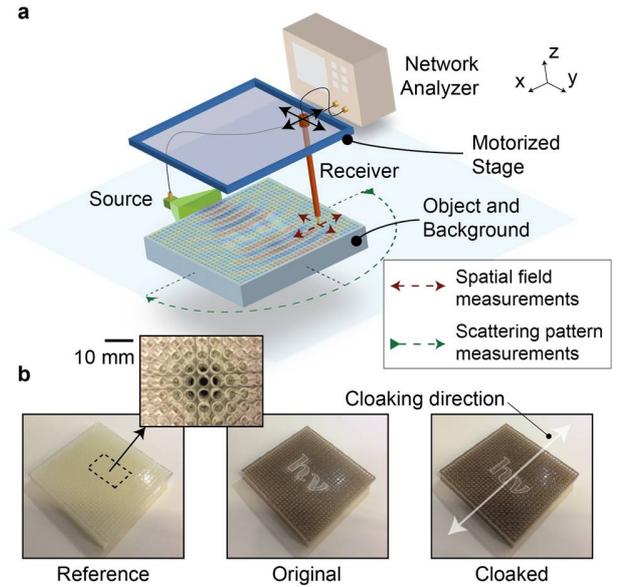

**Fig. 3.** (a) Schematic illustration of the experimental setup. (b) Photographic view of the objects and the background that are engraved into clear cast acrylic (Plexiglas) sheets. From left to right: reference background structure without the object, original object and the resultant cloaked object. The white lines outline the object and the inset shows a zoomed portion of the perforated Plexiglas.

To quantitatively compare the spectrally resolved scattering pattern of the original and the cloaked object, the electric field has been measured rotationally around all three structures with a distance of 135.0 mm with respect to the center of the structures (as indicated by green dashed lines in Fig. 3(a)). To account only for the scattered waves induced by the object, the scattering caused by the diffraction effect of the finite slab structure should be excluded. For this, the scattered electric field values of the reference structure has been subtracted from that of the cloaked and original structures and the absolute values of the differences were taken. Furthermore, the results were normalized by the field spectra obtained without the structures.

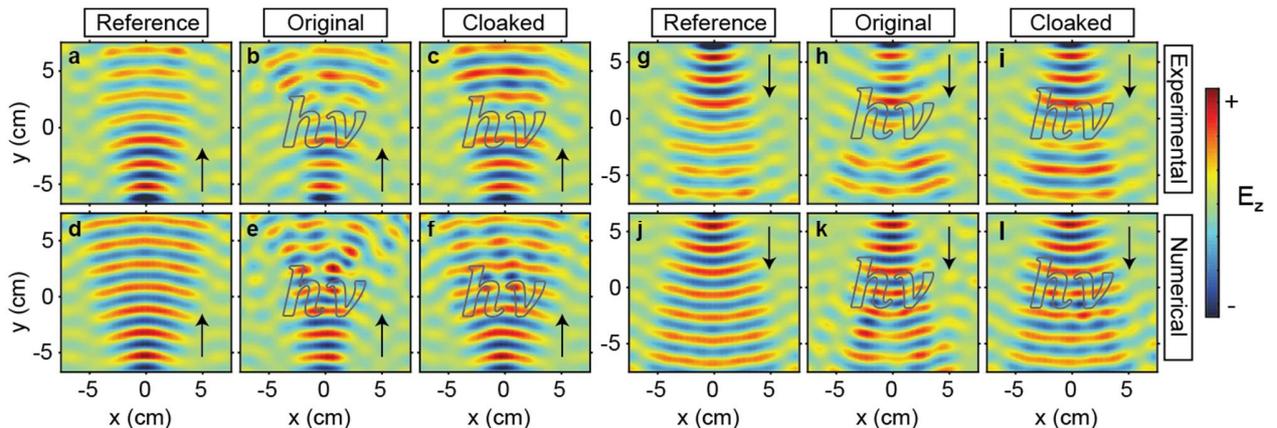

**Fig. 4.** Measured and calculated field distributions for wave propagation in the +y (left) and –y (right) directions at the operational frequency 12.0 GHz. The first row shows the experimental results for the reference background (a,g), original object (b,h) and modified cloaked object (c,i) while the corresponding numerical simulations are depicted in the second row. Similar field measurements for the +x and –x directions, for which the modified object does not exhibit invisibility, and at the operational frequencies 10.5 GHz, and 12.9 GHz are provided in Figs. S3, S4 and S5, respectively (see Supplement 1, Section 3). The black arrows reveal the direction of wave propagation, while the grey lines outline the boundaries of the "hv" shaped object. Moreover, the time evolutions of the measured field profiles are provided in Visualization 1.

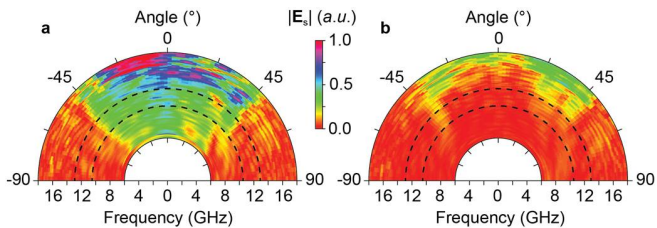

**Fig. 5.** Angular scattered field ($E_s$) amplitude distributions for the (a) original object and (b) the modified cloaked object. Wave propagation is in the +$y$ direction. The scattering angle is defined with respect to the vertical $y$-axis in clockwise direction. The dashed black lines indicate the frequency range for which the object is designed to be cloaked. Similar scattering plots for other incidence directions are provided in Fig. S5 (Supplement 1, Section 4).

Figures 5(a) and 5(b) show the obtained scattering plots for the original and the cloaked objects, respectively; for wave propagation in the +$y$ direction. It can be inferred from these figures that the scattered amplitude decreases substantially in the designed cloaking frequency interval (indicated by black dashed lines) for the cloaked object, as expected. It further follows from these figures that the scattering is not perfectly diminished, which we attribute to the fabrication disorders and to the coupling effect of the receiver antenna with the scattered field. One interesting fact to notice is that for frequencies lower than the designed cloaking frequency interval, the scattered amplitude remains also near zero. We attribute this observation to the fact that the forward-scattering Fourier components along constant wavevector circles, corresponding to lower frequencies, are already included in the filtered wavevector region. In other words, as these constant wavevector circles have a smaller radii, the tangential overlap at the forward-scattering Fourier region between them and the filtered region becomes larger, as compared to high-frequency constant wavevector circles. In particular, this makes our proposed method superior to previously demonstrated cloaking procedures, where the cloaking performance is rapidly degraded away from the central frequency or even drastically enhanced scatterings are observed in nearby regions of the operational bandwidth [21]. Similar scattering plots for other propagation directions have been prepared in Fig. S5 (Supplement 1, Section 4), where an analogous scattering suppression can be observed for wave propagation in the opposite (–$y$) direction. Furthermore, to allow also for an experimental observation of the phase-front response, a broadband scanning of the electric field was performed at the back face of the structures. The experimental setup and the results of these measurements are provided in Figs. S6 and S7, respectively (Supplement 1, Section 5), where one can similarly observe the scattering suppressions and wavefront reconstructions in the designated frequency range.

In this manuscript, we have provided a general description on how to achieve a self-cloaking effect for a given object with arbitrary shape and dimensions, by merely using isotropic and non-magnetic dielectric materials. We demonstrated the feasibility of the proposed method by directly observing the field profile along the structures in an experimental analyses conducted at the microwave regime. We further provided experimentally measured scattering patterns to quantitatively evidence the suppression of the scattered waves in the designed propagation directions and within the specified frequency range. Since the proposed method is independent from spatial dimensions, the same procedure can be applied to realize the proposed invisibility-based cloaking effect at the micro- or nano- scale, as such fabrication methods are already matured owing to the extensive research effort in the field of transformation optics [22-24]. Moreover, in principle, the self-cloaking effect can be also extended to operate under broad-angle incidence (or even omnidirectionally), if the scattered waves are also eliminated for such a broad incidence angle range. The presented work provides a new perspective in the field of optical cloaking, and may be especially valuable for "sensor cloaking" [16] or "low-interference communication" [25] applications where wave penetration into the object is essential instead of wave rerouting around it.

**Acknowledgements.** NATO SPS research grant No: 985048. Spanish Ministerio de Ciencia e Innovación. European Union FEDER project: FIS2015-65998-C2-1-P. Turkish Academy of Sciences.

(†) Online link to supplementary video 1: http://bit.ly/2w55dvE

# All-dielectric self-cloaked structures: supplementary material

**Zeki Hayran,**[1,*] **Ramon Herrero,**[2] **Muriel Botey,**[2] **Hamza Kurt,**[1] and **Kestutis Staliunas** [2,3]

[1] *Nanophotonics Research Laboratory, TOBB University of Economics and Technology, Department of Electrical and Electronics Engineering, Ankara, Turkey*
[2] *Departament de Física, Universitat Politècnica de Catalunya (UPC), Barcelona, Spain*
[3] *Institució Catalana de Recercai Estudis Avançats (ICREA), Barcelona, Spain*

*\*Corresponding author: zekihayran@etu.edu.tr*

---

### 1. Calculation of the effective index and the fabrication of the acrylic slabs

To experimentally realize the graded index distributions in Figs. 2(a) and 2(c) (in the manuscript), we employed the well-known effective medium theory (EMT) to mimic the homogeneous index profile with discrete elementary unit cells. In this regard, the index profiles were divided into 30 divisions/unit cells with spatial dimensions of 5.0 mm in both the *x*- and *y*- directions, to ensure that the lattice constant would be considerably smaller than the operational wavelength. The overall dimensions (150.0 mm x 150.0 mm) of the cloak, on the other hand, is determined so that a complete field scanning is feasible with our experimental setup. The average refractive index for each unit cell has been then obtained by calculating the arithmetic mean of the refractive index inside the unit cell boundaries. On the other hand, to determine the unit cell filling ratio relation with respect to the average index, we performed a series of numerical finite-difference time-domain (FDTD) numerical calculations. Here, we used clear cast acrylic sheets which have a bulk refractive index equal to 1.59. To calculate the effective refractive indices for various hole radii and slab thickness of 24.0 mm, the spatial wavelength of the propagating mode inside a periodic structure with the respective constant hole size and the same slab thickness has been numerically obtained. The obtained spatial wavelength has been then divided to the spatial wavelength in free space to determine the effective index for the slab with the respective hole size. The results are indicated as red circles in Fig. S1, while the solid blue line shows the fitted curve obtained in the analytical form of

$$n_{\mathit{eff}} = \sqrt{(-0.1599 mm^{-2})r^2 - (0.0616 mm^{-1})r + 2.3320},$$
**(S1)**

where r is the radius of the hole. Moreover, for a comparison, the Maxwell-Garnett model2 for the corresponding two-dimensional (2D) structures (having infinite slab thickness) have been superimposed in Fig. S1 as a dashed green line. As it follows from Fig. S1, the realizable effective index for a 24.0 mm thick acrylic slab ranges between 1.09 and 1.50. In light of this information, the index profiles given in Fig. 2 (in the manuscript) have been adjusted such that the maximum and minimum average refractive indices does not exceed these limits.

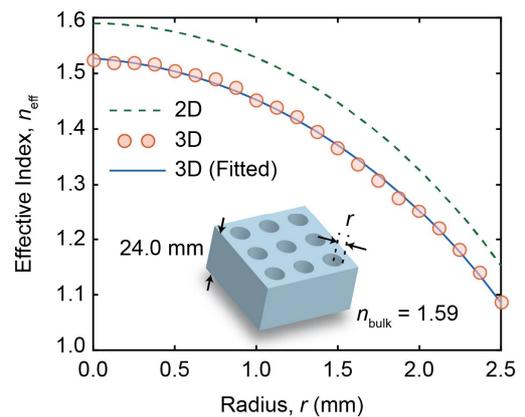

**Fig. S1.** Effective index variation for various hole radii, superimposed for the 2D (slab with an infinite thickness) and the 3D (with a thickness equal to 24.0 mm) case.

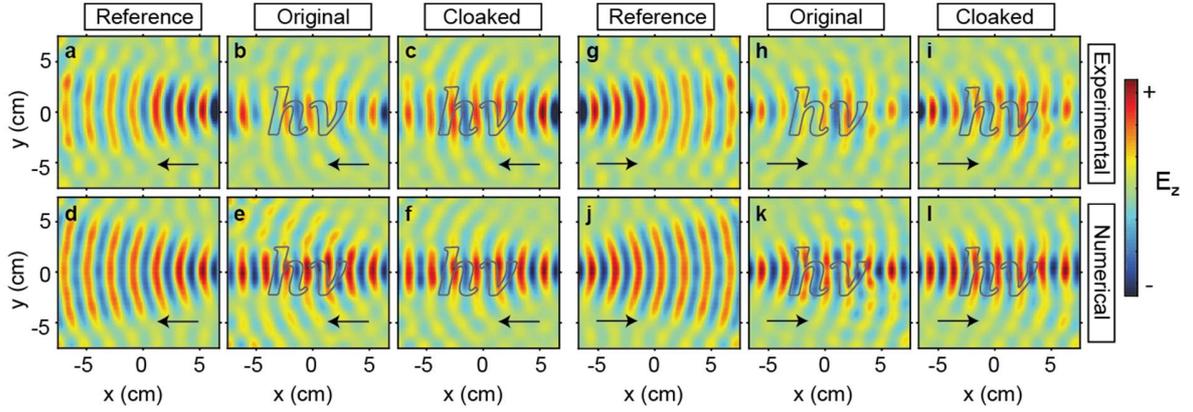

**Fig. S2.** Measured and calculated field distributions for wave propagation in the -*x* (left) and +*x* (right) directions at the operational frequency 12.0 GHz. The first row shows the experimental results for the reference background (a,g), original object (b,h) and modified cloaked object (c,i) while the corresponding numerical simulations are depicted in the second row. The black arrows reveal the directions of wave propagation, while the grey lines outline the boundaries of the "*hv*" shaped object.

Using Eq. 1, the discretized index profiles were then converted into a bulk material with unit cells containing the calculated hole sizes. The resulting structure was fabricated into clear cast acrylic sheets using a laser engraving machine (Superstar C*X*-1290) which has a engraving position accuracy of 50 µm. Due to the lack of 24.0 mm thick acrylic sheets, we instead employed and perforated three 8.0 mm thick sheets, which were then stacked on top of each other. Special care was taken to perfectly align the holes and to avoid any air gap between the sheets. The photographic illustration of the fabricated structures are given in Fig. 3(a) (in the manuscript).

## 2. Experimental measurement technique

The setup setup was placed in an anechoic camber constructed by microwave absorbers, to avoid undesired environmental reflections. The horn antenna has an aperture size of 41.4 mm and 32.3 mm in the *x*- and *z*- directions, respectively; and has been placed 10.0 mm in front of the structure to inject electromagnetic waves with electric field polarized in the *z*- direction. On the other hand, the receiver monopole antenna has been kept parallel to the *z*- axis throughout the experiments to ensure the polarization match with the transmitter horn antenna. Since we are not able to measure the field directly inside the structure, we instead obtain the field by interacting the tip of the monopole antenna with the evanescent tails of the field that propagates inside the structure. In this regard, the tip of the monopole antenna was placed 1.0 mm above the top surface of the structure and a total area of 150.0 mm and 142.0 mm in the *x*- and *y*- directions, respectively, have been scanned with a 1.0 mm spatial resolution via the motorized stage, as indicated by red dashed lines in Fig. 3 (in the manuscript) (due to limitations of the motorized stage, a beginning part of 8.0 mm in the *y*- direction could not be measured). In this way, the three fabricated structures (background reference structure without object, original object and the cloaked object, see Fig. 3(b)) have been illuminated from all four directions (+*x*, +*y*, -*x*, -*y*) and the corresponding electric field distributions along the top surface of the structures have been measured.

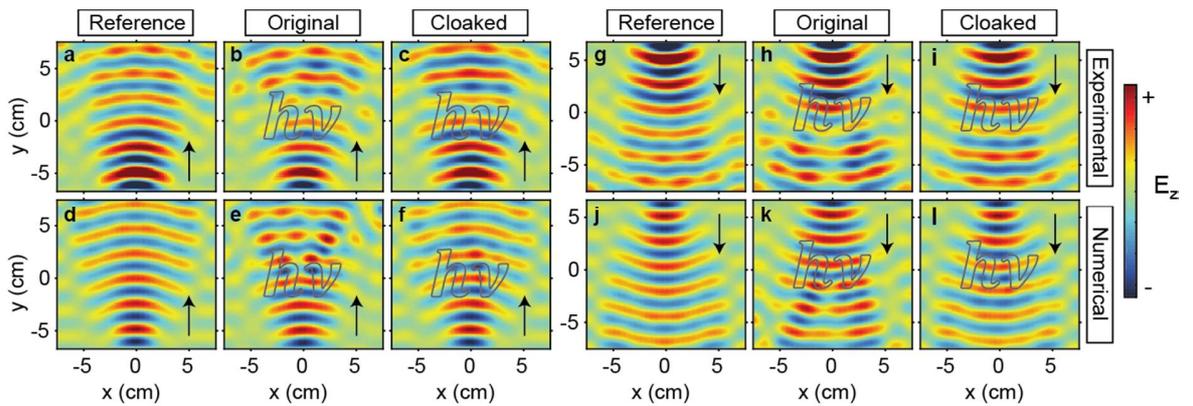

**Fig. S3.** Measured and calculated field distributions at the operational frequency 10.5 GHz. The first row shows the experimental measurements, while the second row shows the corresponding numerical calculations. The field profiles were obtained along the (a,d,g,j) reference, (b,e,h,k) original and the (c,f,i,l) cloaked structures. The black arrows reveal the direction of wave propagation, while the grey lines outline the boundaries of the "*hv*" shaped object.

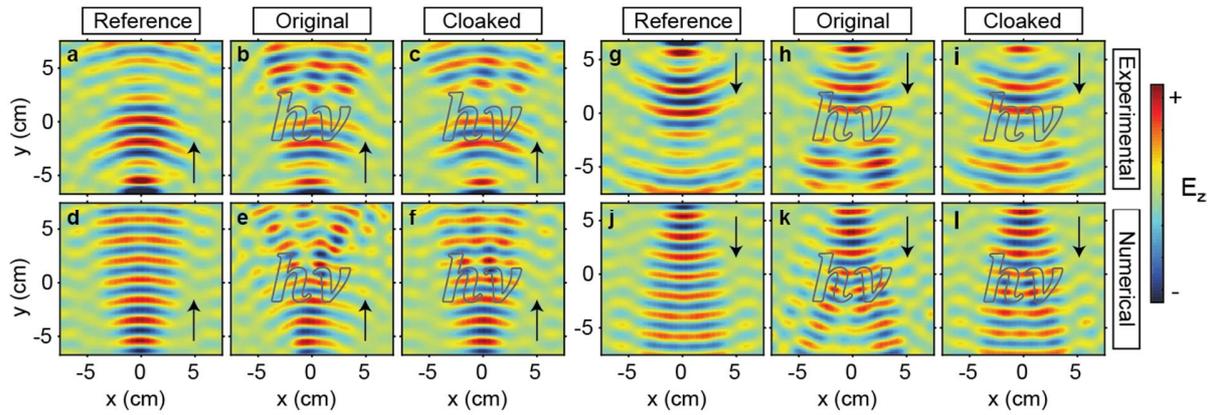

**Fig. S4.** Similar measurements and calculations as in Fig. S3 obtained at the operational frequency 12.9 GHz. The black arrows reveal the direction of wave propagation, while the grey lines outline the boundaries of the "*hv*" shaped object.

### 3. Measured spatial field distributions

To verify that the structure does not exhibit invisibility outside the designed +*y* and –*y* directions, the spatial field profiles for wave propagation in the +*x* and –*x* directions were measured. The results are collectively presented in Fig. S2 with their corresponding numerical counterparts. As it appears from these results, the field profile along the cloaked object does not mimic the reference field profile and differs only slightly from the field profile of the original object. The reason for the discrepancy between the field profiles of the original and the cloaked objects is that a part of the filtered wavevector region (in the ±*y* direction) overlap with the constant wavevector circles for the ±*x* directions. Hence, a portion of the scattering waves (around the ±*x* directions) would be affected by the wavevector filtering (in the ±*y* direction).

The measured field profiles for wave propagation in the ±*y* directions at the frequencies 10.5 GHz and 12.9 GHz have been provided in Figs. S3 and S4, respectively; to evidence the broadband operation designed by the wavevector filtering in Fig. 2 (in the manuscript). Similar as in Fig. 4 (in the manuscript), Figs. S3 and S4 demonstrate that the cloaked object recovers the wavefront shape of the incident wave as in the objectless reference structure, whereas the original object causes various light scatterings and wave deformations. Important to notice is that the as the operational frequency increases, the scattered waves become more dominant as predicted by the numerical scattering pattern analyses presented in Figs. 5 (in the manuscript) and S5.

### 4. Scattered intensity plots

Further scattering analyses have been performed for wave propagation in the –*y*, +*x* and the –*x* directions and have been presented in Figs. S5 (a, b), S5 (c, d) and S5 (e, f), respectively. As it follows from these figures, in the designed frequency range (indicated by black dashed lines), the scattered waves diminishes substantially for wave propagation in the –*y* direction (Fig. S5 (c, d)). On the other hand, no scattering reduction occurs for the +*x* (Fig. S5 (a, b)) and the –*x* (Fig. S5 (e, f)) directions, as expected. Interesting to note is that the scattered intensity plots undergo a change when the invisibility effect is enabled for the –*x* and +*x* directions. As was also discussed in the previous section, we attribute this change to the partial overlap between the constant wavevector circles in the ±*x* directions and the filtered wavevector region in the ±*y* directions.

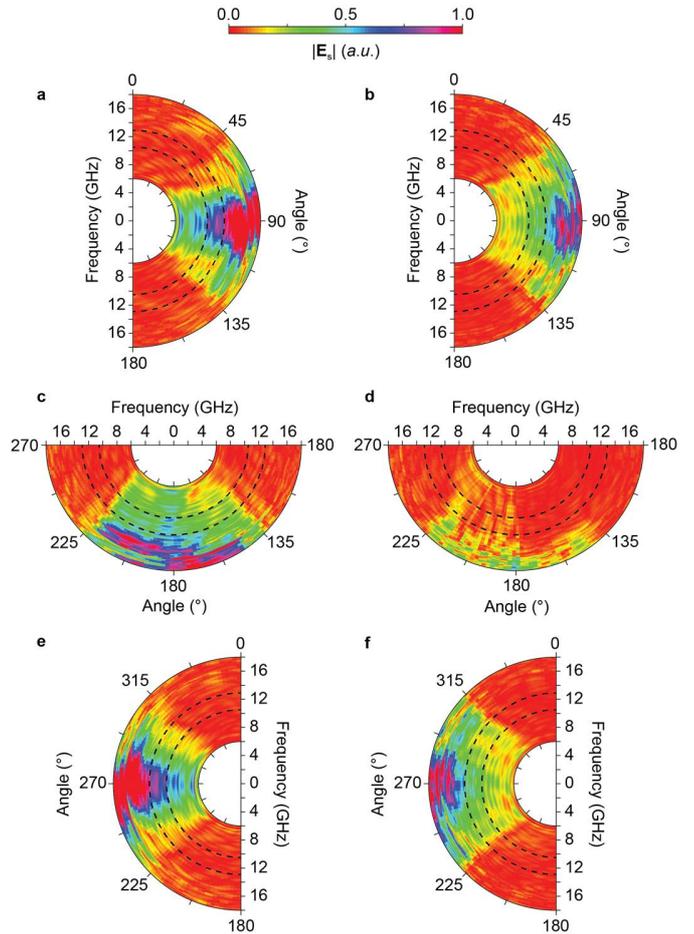

**Fig. S5.** Pattern of scattered intensity for (a, c, e) the original object and (b, d, f) the cloaked object. Wave propagation is in the (a, b) -*y*, (c, d) +*x* and the (e, f) -*x* directions. The scattering angle is defined with respect to the vertical *y*-axis in clockwise direction. The dashed black lines indicate the frequency interval at which the object is designed to be cloaked.

## 5. Measurement of the electric field spectra

To experimentally obtain a broadband phase-front response, the electric field was measured with a distance of 10.0 mm with respect to the back face of the structure, along the center of the structural height in the z- direction (see Fig. S6). The monopole antenna was stepped in the x- (for wave propagation in the ±y directions) or the y- (for wave propagation in the ±x directions) direction with 1.00 mm steps. Figure S7 shows the measured field spectra for all three structures in the frequency interval of 6-18 GHz. Similar to our previous analyses, the field spectra obtained from the reference structure (Fig. S7(a)) allows us to verify the cloaking operation of the cloaked structures. In this regard, when comparing the field spectra of the cloaked structures (Figs. S7(c) and S7(g)), one can deduce that the scatterings observed in the original structure (Figs. S7(b) and S7(f)) have been reduced. While the resemblance between the reference field spectra (Fig. S7(a)) and the cloaked field spectra in the –y direction (Fig. S7(g)) is remarkable, a slight field distortion can be observed in the cloaked field spectra in the +y direction (Fig. S7(c)). We attribute this discrepancy to potential field disturbances caused by the measurement technique and to possible imperfect alignments of the experimental setup. Moreover, the field spectra for wave propagation in the –x and +x direction have been also measured and provided in Figs. S7(d) and S7(h) for the original structure and in Figs. S7(e) and S7(i) for the cloaked structure. In accordance with our previous analyses, one can see from these figures that no cloaking effect exists for these directions.

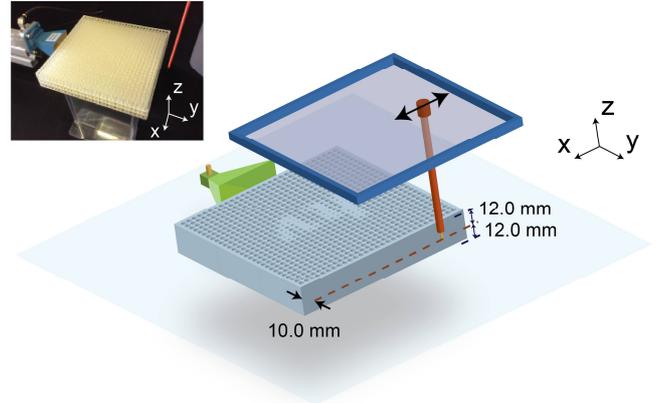

**Fig. S6.** Experimental setup for the field spectra measurement is shown. The monopole antenna is stepped in the lateral direction 10.0 mm behind the structure. The upper left inset depicts the photographic view of the experimental setup.

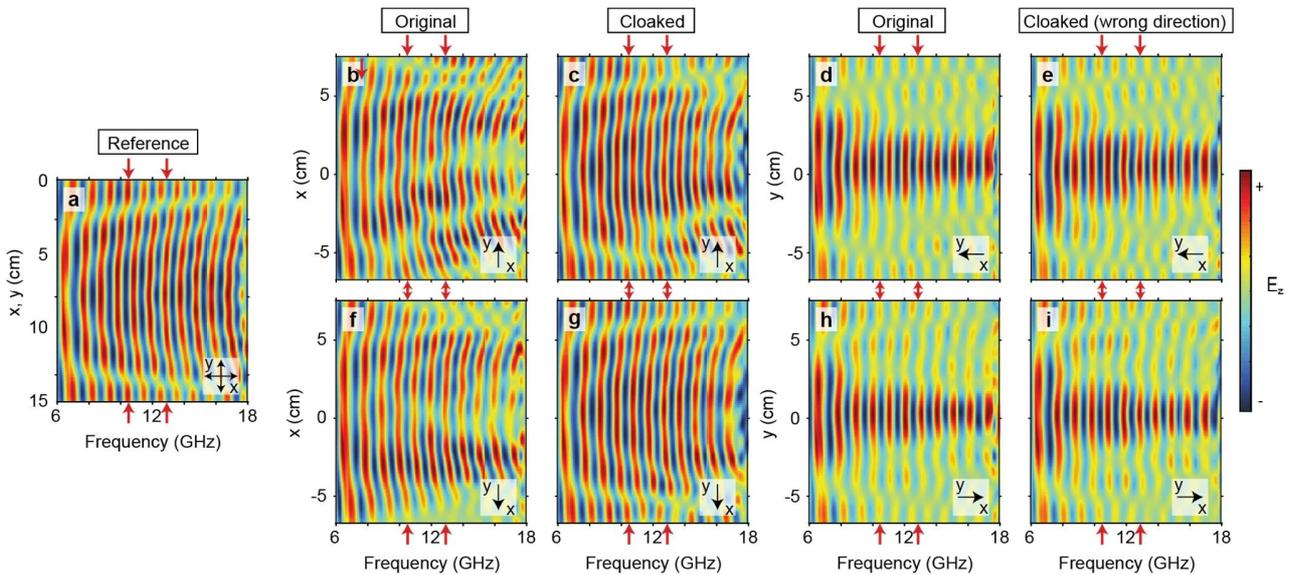

**Fig. S7.** Measured electric field spectra are given. Measurements were taken from the (a) reference, (b, d, f, h) original and the (c, e, g, i) cloaked structure. The lower right insets show the direction of wave propagation, while the red arrows on the horizontal axes delimit the operational bandwidth.